\documentclass[12pt]{article}
\usepackage{epsf}
\usepackage{alltt}
\usepackage{amsmath}
\usepackage{epsfig}
\def\largelinestretch{\renewcommand{\baselinestretch}{1.1}}
\largelinestretch\small\normalsize

\title{
\vspace*{-15mm}
{\normalsize 
\begin{tabular}[t]{ll}
ZEUS  \hspace{0.5cm} 06--10 &\\  
\today & 
\end{tabular}
\hfill {} \\}
\vspace*{15mm}
{\bf  Improvement of the Kalman Filter Fit of the ZEUS experiment}
}

\author{
       {\it Vincent Roberfroid}\footnote{
E-mail: vincent.roberfroid@desy.de 
}\\
        DESY }
\date{}

\begin{document}
\maketitle
\begin{abstract}
At the center of the ZEUS detector situated on the collider HERA at DESY in Hamburg, a Micro Vertex Detector has been installed in 2000/01. Situated close to the beam line, this subdetector is designed to measure precisely the tracks near the interaction point (the vertex).

A package called KFFIT is designated to perform the track fit of the Micro Vertex Detector hits using a Kalman Filter. 

This paper is a report of all improvements done in this package.
After a brief description of the algorithm, the improvements of KFFIT will be presented comparing the last version (v10 released in December 2006) to the initial one (v1 released in December 2005 (=v2005a.1)).  
\end{abstract}

 \section{Introduction}
Before the luminosity upgrade of HERA in 2000/01, the tracking system was made of the Central, 
the Forward and the Rear Tracking Detector (CTD, FTD and RTD) and the Transition Radiation Detector (TRD).
During the upgrade of HERA, a micro-vertex detector (MVD) 
and a straw tube tracker (STT), which replaced the TRD, have been introduced in the ZEUS detector and added to the tracking system.
The ZEUS tracking algorithms had in consequence to be enhanced to include the 
additional information.
In particular, a new package called KFFIT has been written for adding the MVD 
hit positions into the track fit. 
This algorithm is extensively described in reference \cite{MAD04}.
In 2005, a complete description of the MVD geometry has been included in KFFIT allowing a proper
calculation of the covariance matrices.
Nevertheless, new ideas have been proposed to improve substantially the momentum 
resolution. These ideas consisted essentially in using proper input tracks for KFFIT.
During the modification of the algorithm it has been found that other modifications could 
improve further the momentum resolution, the computing time and the 
calculation of the covariance matrices.

All these modifications will be described in this note.
But, to be able to understand the improvements realized, a brief description of 
the tracking chain and a short description of the algorithm itself will be done.
Then a section describing the new modifications will be followed by a comparison 
of the various resolutions obtained with the latest version of the algorithm (v10 : released in December 2006) 
and the initial one (v1 : released in December 2005).

\section{KFFIT and the tracking chain}
\subsection{The tracking reconstruction chain} \label{subsec:track-chain}
\begin{figure}[hbtp]
\centering
\includegraphics[height=6.0cm]{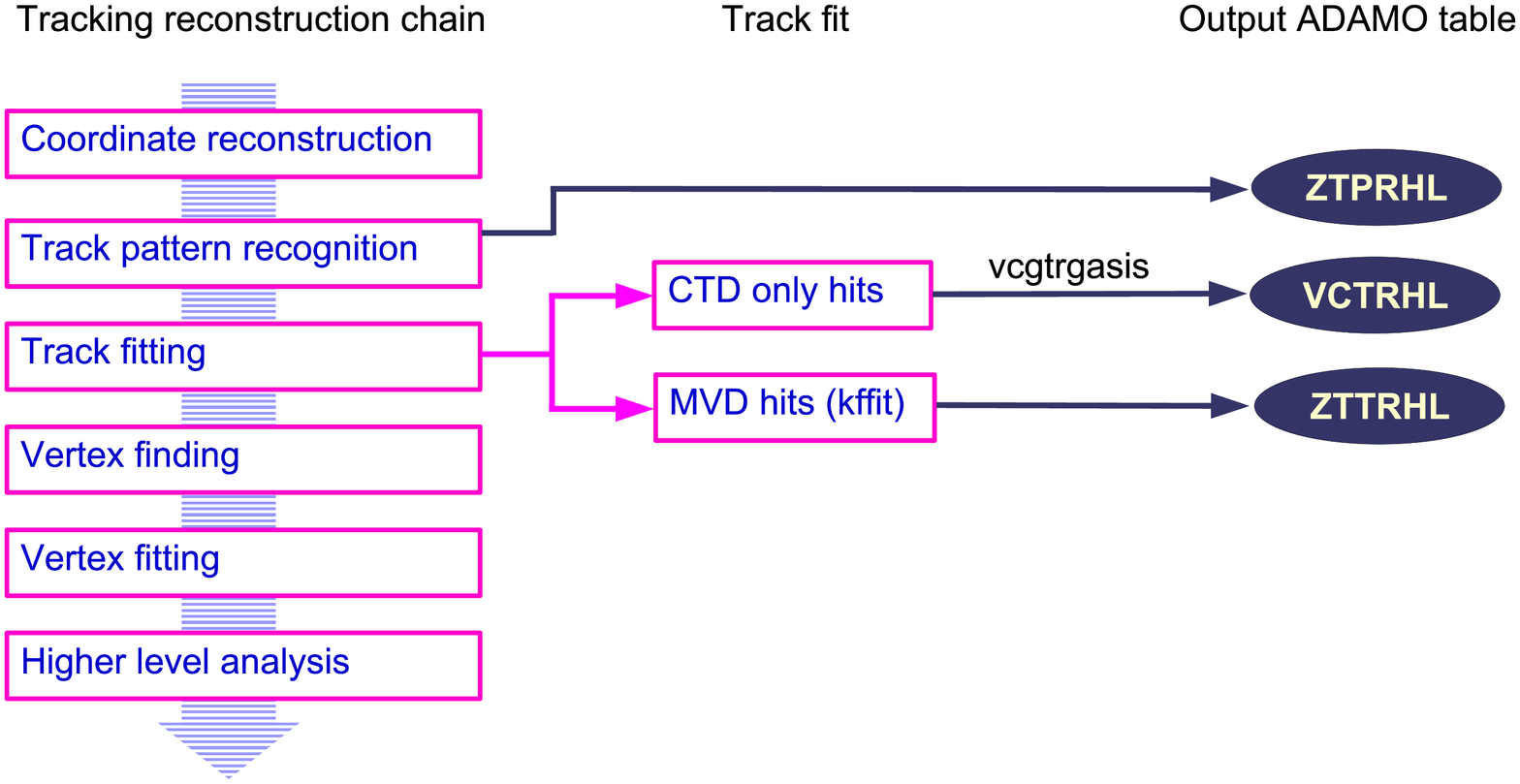}
\caption{The track reconstruction chain, the 2 separated fitting packages and the available output ADAMO tables.}
\label{fig:track-chain}
\end{figure}

The tracking reconstruction chain is composed of different steps displayed on figure \ref{fig:track-chain} :
\begin{itemize}
\item[$\bullet$] The coordinates of the hits in the CTD, MVD and STT detectors are reconstructed.
\item[$\bullet$] The pattern recognition combines hits together to construct the ZTPRHL tracks (in the VCRECON package \cite{HAR98}).
\item[$\bullet$] The CTD hits are globally fitted by an helix and the output tracks can be stored in the ZTPRHL table with the call of the subroutine VCGTRGASIS (not by default).
\item[$\bullet$] The MVD hits are added to the fit by the use of a Kalman Filter (KFFIT) \cite{MAD04}. 
\item[$\bullet$] Finally the vertices are identified and fitted (either by KFVTX, second part of the KFRECON package, or by the new package: DAF).
\end{itemize}
The VCTRHL and ZTPRHL tracks are computed by the package VCRECON \cite{HAR98}.

The ZTPRHL ADAMO tables contain output tracks from the pattern recognition. Each row in this table contains a link to the MVD and the STT hits (MVRECC)
associated to the track.
The track parameters and covariance matrices stored in the ZTPRHL table have been obtained using CTD+MVD+STT hits with material corrections applied until the interaction point.
However they don't come from a rigorous track fit and are thus not as precise as the ZTTRHL tracks.

The VCTRHL ADAMO tables are filled by two distinct routines : VCGETREG which fills ZTPRHL tracks which are associated to a vertex (these tracks are also called REGULAR tracks)
and VCGTRGASIS which fills the fitted CTD only tracks (we will call these tracks VCTRHL CTD-only tracks).
These VCTRHL CTD-only tracks are obtained from a rigorous global fit of all the CTD hits with material corrections until the interaction point\footnote{The material correction between the CTD inner wall and the interaction point is however done with inappropriate geometry assumptions in VCRECON.}.

The ZTTRHL tracks are computed by KFFIT, the first part of the KFRECON package which will be described here below\footnote{The ZTTRHL table is extensively describes in Annexe 1 at the end of this note.}.

\subsection{The initial form of the KFFIT algorithm}
KFFIT is a Kalman Filter which adds step by step the MVD hits to a track from the inner wall of the CTD to the interaction point through the three cylinder of the MVD
and calculate for each step the parameters $\vec{a}$ and the covariance matrix $\boldsymbol{C}$.\\
\subsubsection{Parameters and covariance matrix}
\begin{figure}[hbtp]
\centering
\includegraphics[height=4.0cm]{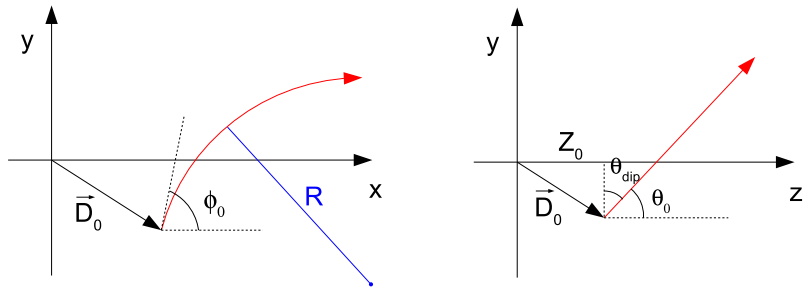}
\caption{Track in an homogeneous magnetic field projected on the xy and the z planes. The track parameters $Q/R$, $cot(\theta_{dip})$, $\phi_0$, $D_0$ and $Z_0$ used in KFFIT are shown.}
\label{fig:param}
\end{figure}

In KFFIT, the 5 parameters which define a track are : 
\begin{equation}
\vec{a} = (Q/R, cot(\theta_{dip}), \phi_0, D_0, Z_0)
\end{equation}
All these parameters are explained graphically in figure \ref{fig:param}.

Q is 1 if the bend of the track is anticlockwise and -1 if the bend of the track is clockwise in the xy plane. 
R is the radius of the helix.
$\theta_{dip} = 90-\theta_0$ where $\theta_0$ is the angle of the track with respect 
to the z axis at the point of closest approach to the origin.
$\phi_0$ is the angle of the track tangent at the point of closest approach with respect to the x axis.
$D_0$ is the length of the vector $\vec{D}_0$, the point of closest approach and
$Z_0$ is its z coordinate.
\\
Moreover a 5x5 covariance matrix $\boldsymbol{C}$ is associated to these parameters giving as diagonal elements 
the variance of each parameters.

In the three following paragraphs, the Kalman Filter procedure used to calculate the parameters and the covariance matrix is described.

\subsubsection{Initialisation of the Filter} \label{subsec:initKF}
In version v1 (v2005a.1), the input track parameters $\vec{a_0}$ and covariance matrix $\boldsymbol{C}_0$ 
used in KFFIT was extracted from the ZTPRHL table. 
These tracks contain already the STT+CTD+MVD hits. 
So using these tracks as input for KFFIT leads to the problem
that the MVD hits are double counted. That forces the tracks to pass through the MVD hits and so can modify
their bending and in consequence the estimated momenta.

Moreover, these tracks were not obtained from a rigorous fit and the associated covariance matrices are not 
consistent as they depend on the method used to perform the pattern recognition 
(3 different methods are envisaged for each tracks \cite{HAR98}).

In consequence, the initial track parameters and covariance matrices used as input of KFFIT have been modified as described in section \ref{subsec:input-track}.

\subsubsection{Energy loss and multiple scattering} \label{subsec:eloss}
At each step $k$ of the Kalman Filter, the multiple scattering effects in the dead material crossed
by the particle between the step $k-1$ and the current step $k$ is calculated and added to 
the covariance matrix $\boldsymbol{C}_k$ :
\begin{equation}
\boldsymbol{C}'_k=\boldsymbol{C}_k+\boldsymbol{M}_k
\end{equation} 
$\boldsymbol{M}_k$ is a 5x5 matrix obtained by calculating the contribution of the angular 
uncertainty induced by the multiple scattering on all the track parameters. 
This matrix depends on the radiation length of the scattering medium \cite{AVE92}.
\\
Then the energy loss and its impact on the track parameters is then calculated with the following procedure :
\begin{itemize}
\item The transverse momentum $p_T$ is calculated converting the curvature ($Q/R$) using the magnetic field 
calculated at the origin (0,0,0) and so assuming that the magnetic field is constant in all the MVD.
\item The momentum loss in dead material between the step $k-1$ and $k$, $\Delta p_T$, is calculated using the Bethe-Bloch formula assuming that all 
particles are pions (mass=0.1396 GeV/c$^2$). 
Moreover, since information on materials in KFFIT does not include the densities and the atomic numbers, but only the radiation thickness, 
it has been necessary to assume that all material is composed of Silicon (Z=14, $\rho=2.33$ g/cm$^3$) only. 
\item The new curvature is calculated : $W'=W*p_T/(p_T+\Delta p_T)$ where $W=Q/R$.
\item The average position $\vec{r}$ in the dead material is calculated.
\item The new track is calculated assuming that this track must have the same direction than the initial track
at $\vec{r}$ but with a curvature $W'$.
\end{itemize}
So, in contrast to a global fit, the track parameters and the covariance matrices depend on the position in the MVD.
Using a Kalman Filter, KFFIT does not assume that the track follows a single helix, but a succession of helix. 

\subsubsection{Addition of an MVD hit}
All the MVD hits coming from the pattern recognition (MVRECC hits linked to the ZTPRHL track) are added to the track, starting 
from the outer to the inner cylinder of the MVD.
Each time a MVD hit is added to the track at a given step $k$, the parameters and the covariance
matrix are updated by the following equations \cite{AVE92} :
\begin{eqnarray}
\boldsymbol{C}_k & = &\boldsymbol{C}'_{k-1} (\boldsymbol{1}+\vec{g}_{k-1}^T\vec{g}_{k-1} \boldsymbol{C}'_{k-1} \sigma_{k})^{-1}  \\
\vec{a}_{k} & = & \vec{a}_{k-1}+ \boldsymbol{C}_k \vec{g}_{k-1} \sigma_{k} \Delta x
\end{eqnarray}
where $\vec{g}$ is the "derivative vector" :
\begin{equation}
\vec{g}_{k-1} = \frac{\partial f(\vec{a}_{k-1})}{\partial \vec{a}}
\end{equation}
where $f$: $\boldsymbol{R}^5\rightarrow\boldsymbol{R}$ is a function giving the expected local position 
on the sensor perpendicularly to the strips.\\
$\Delta x$ is the difference between the MVD hit position and the intersection position of the extrapolated track with the sensor.
$\sigma_{k}$ is the resolution of the $k$\_th MVD hit.
This resolution depends on the angle of the track with respect to the normal of the sensor.
A track crossing a sensor with larger angle induces charges on more strips and so a larger cluster size.
The parametrization used in KFFIT is the same as the one used in the Monte Carlo reconstruction and 
has been determined using an electron test beam at DESY \cite{MAD04} :
\begin{eqnarray}
\sigma_k = 10.13+22*\theta^2+\frac{1}{cos(\theta)} \\
\sigma_k' = \sqrt{\sigma_k^2+40^2}
\end{eqnarray}

where $\theta$ is the angle (in radian) of the track with respect to the normal of the sensor.
$\sigma_k'$ is the resolution of the MVD hit (in microns) for real data in which a contribution of 40 $\mu m$ has been added to take into account misalignment of the MVD.

\section{Improvement of KFFIT}
\subsection{Input track}
\subsubsection{ADAMO table to be used} \label{subsec:input-track}
The ideal initial track parameters $\vec{a_0}$ and covariance matrices $\boldsymbol{C}_0$ 
would be those obtained using only the CTD hits with material correction done only until the outer radius of the MVD.
But unfortunately, an ADAMO table containing such information does not exist. 
However, it is possible to calculated those tracks using the VCTRHL CTD-only tracks (see paragraph \ref{subsec:track-chain}).
Unfortunately these tracks are computed at the vertex and not at the outer radius of the MVD.
So, the covariance matrices associated with these tracks contain already multiple scattering effects between the
CTD and the interaction point. In consequence, this multiple scattering effect had to be recalculated
and removed from the VCTRHL CTD-only covariance matrices. A new routine has been written by G. Hartner  
for this purpose (called VCUNDOSCAT\footnote{This subroutine is now integrated in KFRECON. 
It is written in fortran and does basically the same job as the subroutine VCSCAT from VCRECON, but removing scattering effects instead of adding them.}). 

\begin{figure}[hbtp]
\centering
\includegraphics[height=10cm,angle=-90]{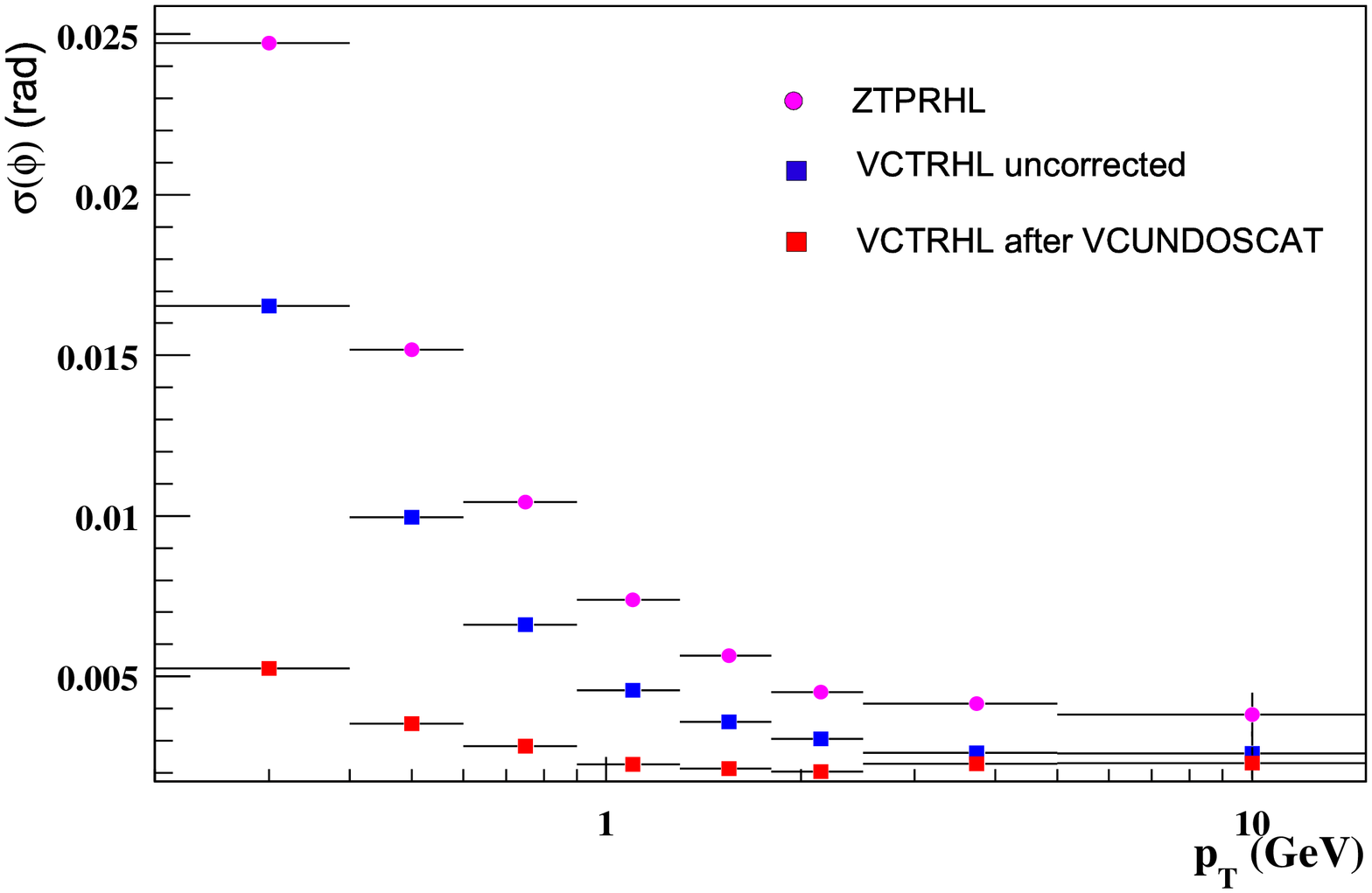}
\caption{Estimated error of the track parameter $\phi$, $\sigma(\phi)$ when using different input tracks : ZTPRHL, VCTRHL (ctd only) uncorrected, VCTRHL (ctd only) after VCUNDOSCAT.}
\label{fig:sigma-phi}
\end{figure}

The figure \ref{fig:sigma-phi} shows the estimated standard deviation $\sigma(\phi)$ of the track parameter $\phi$ (which is simply the square root of one diagonal element 
of the covariance matrix) as a function of the transverse momentum $p_T$ for various
possible input tracks. The full pink circle corresponds to the ZTPRHL tracks, the blue squares corresponds to the VCTRHL tracks without any modifications,
and the red squares corresponds to the VCTRHL after removing multiple scattering effect with VCUNDOSCAT.
It appears, as expected, that the predicted $\phi$ resolutions is more depent on the momentum for the ZTPRHL and for the VCTRHL uncorrected tracks since they include material correction up to the interaction. 
This dependency on the transverse momentum has been considerably reduced thanks to VCUNDOSCAT.
A small dependency remains due to the multiple scattering in the CTD and due to the approximations of the momentum used as input of VCUNDOSCAT as explained here below.

\subsubsection{Input momentum correction for VCUNDOSCAT}
\begin{figure}[hbtp]
\centering
\includegraphics[height=4cm]{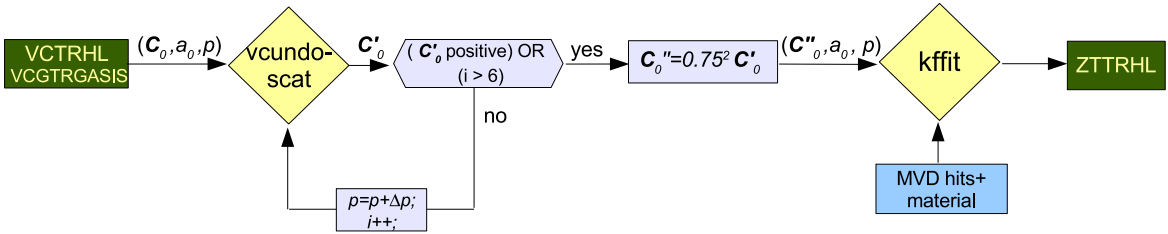}
\caption{Algorithm diagram to produce input tracks for KFFIT.}
\label{fig:input-track}
\end{figure}

The variables used as input for VCUNDOSCAT are the track parameters $\vec{a}_0$ , the covariance matrix 
$\boldsymbol{C}_0$ and the momentum $p$. This last variable should be exactly the same as
the value used initially by the CTD fit package (VCRECON) producing the VCTRHL tracks. 
Unfortunately this variable is not stored in an ADAMO table making a rigorous solution impossible.
So, to a good approximation, the momentum used instead was the momentum stored in the VCTRHL table.
However, in some cases, this VCTRHL momentum can be smaller than the momentum initially used in VCRECON resulting in an 
over-estimation of the multiple scattering matrix $\boldsymbol{M}$. So substracting $\boldsymbol{M}$ from the
covariance matrix $\boldsymbol{C}_0$ can give a non-positive and/or non invertible input covariance matrix for KFFIT.
To solve this problem an iterative procedure was needed as presented on the diagram in the figure \ref{fig:input-track}.

At each iteration, the momentum used as input of VCUNDOSCAT was increased by a value $\Delta p$ equal to the estimated energy loss 
between the CTD and the interaction point. The loop stops when the output matrix $\boldsymbol{C'}_0=\boldsymbol{C}_0-\boldsymbol{M}$ is positive or
when the number of iteration is larger than 6. In this last case, the covariance matrix used as input of KFFIT is 
the uncorrected VCTRHL covariance matrix. 
\begin{figure}[hbtp]
\centering
\includegraphics[height=9cm,angle=-90]{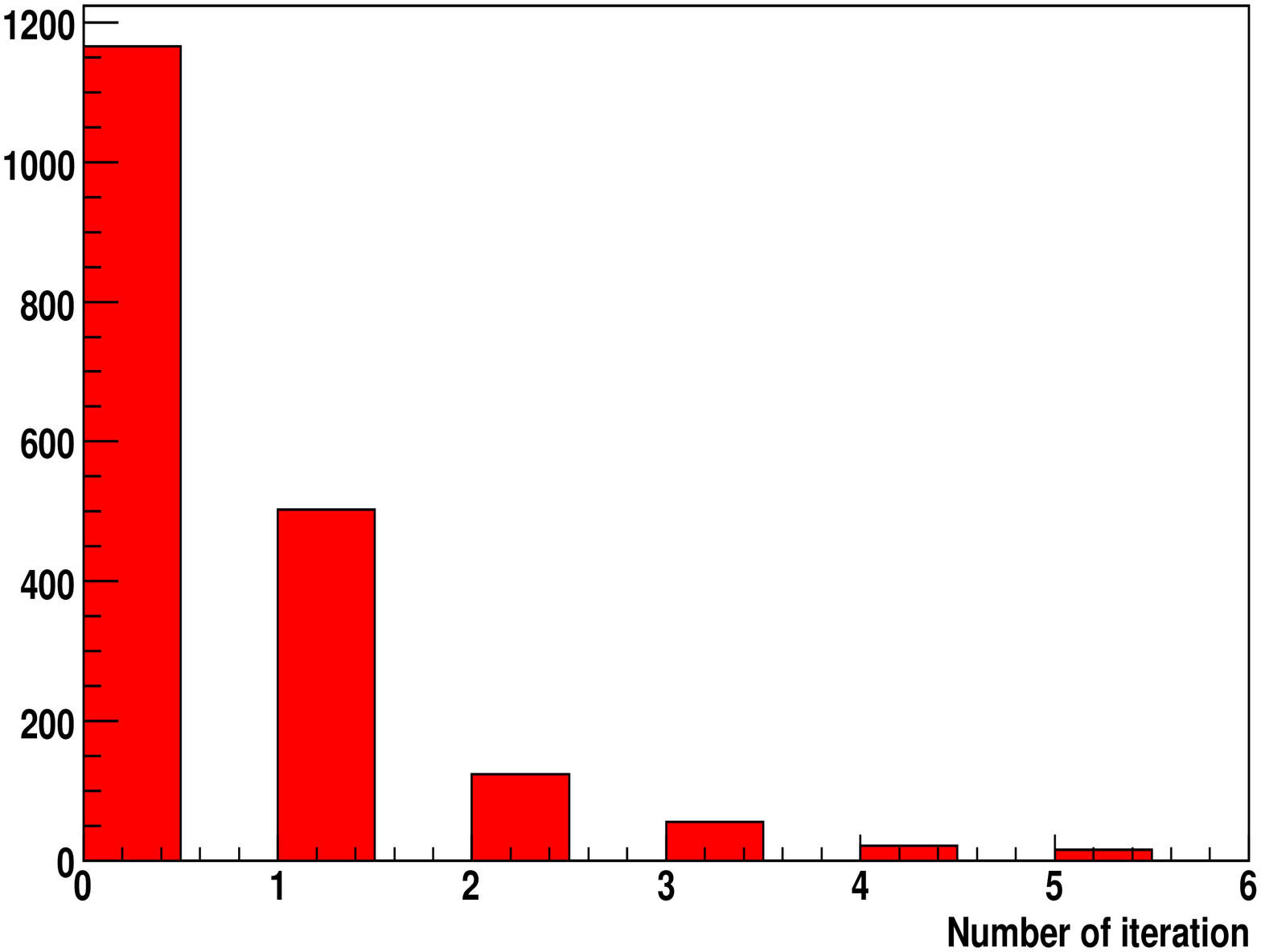}
\caption{Number of iterations needed to produce positive scattering matrix.}
\label{fig:iterations}
\end{figure}
The figure \ref{fig:iterations} shows the number of iterations needed to provide positive covariance matrix using this algorithm.
In most of the cases, no iterations are needed indicating that the the VCTRHL momentum is a good approximation.
It can be seen also that only a few covariance matrices cannot be computed by VCUNDOSCAT within 6 iterations.

\subsubsection{Rescaling of the covariance matrix}
An interesting variable to be studied is the pull of the parameters. The pull of a parameter $a$ is defined by:
\begin{equation}
	pull=\frac{Reconstructed\ parameter - True\ parameter}{Estimated\ error}=\frac{\Delta a_i}{\sigma_{a_i}}
	\label{eq:pull}
\end{equation}
The estimated error $\sigma a_i$ is the square root of the associated diagonal element in the covariance matrix.
\\
With this definition, the standard deviation of the pull $\sigma (pull)$ should be equal to 1 if the estimated error 
is consistent with the Monte Carlo simulation.
However it has been found that the pull of all the input track parameters after VCUNDOSCAT,
were always below 1 at high momentum and around 0.75.

The fact that the pull was below 1 at high momentum cannot come from a wrong estimation of the multiple scattering as 
at high momentum where multiple scattering is negligible. 
Instead, the effect can be attributed to the fact that the CTD hit resolutions have been over-estimated in VCRECON.
This effect is well known and has been used for an optimization of the track-finding efficiency in the pattern recognition.
In consequence, the covariance matrices have been rescaled to take into account this effect (as seen on the diagram Fig. \ref{fig:input-track}).
\subsection{Magnetic field}
As described in paragraph \ref{subsec:eloss}, the magnetic field is used to convert the curvature to the momentum.
This momentum is used in turn to calculate the energy loss and the multiple scattering correction. 
In the initial version of KFFIT the magnetic field used was 
a constant value corresponding to the magnetic field at the origin (0,0,0) which is not appropriate since the field is not perfectly homogeneous. 
This behaviour was changed such that the magnetic field is 
calculated at the position where the conversion curvature to radius is needed\footnote{Using the standard routine GUFLD}.
\\
This modification corrects the final momentum estimation by a factor 0.1 to 0.2 \%.
This is relatively small, but not negligible as we will see in the next section.
\subsection{Momentum bias}
\label{sec:bias}
\begin{figure}[hbtp]
   \centering
   \includegraphics[height=8cm]{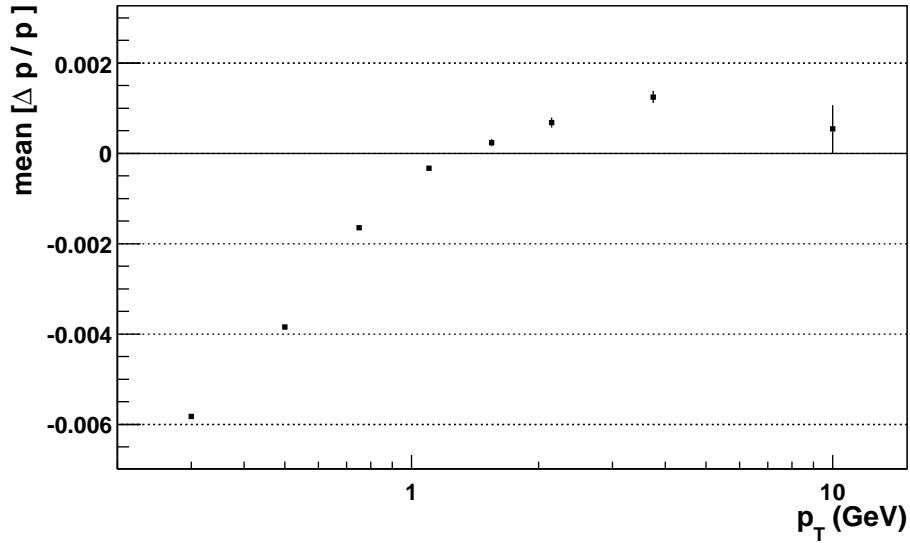}
   \caption{Mean value of the relative momentum difference (reconstructed minus true) as a function of the transverse true momentum.}
   \label{fig:meanDPoverP}
\end{figure}

Some physics analyses, which reconstruct mass distributions of particles, have observed
shifts of some MeV in the mean value of the distribution with respect to the tabulated value.
This shift was coming from a momentum bias in KFFIT.
To illustrate this, the mean difference between the reconstructed momentum and the true value (FMCkin) 
has been plotted as a function of the true transverse momentum in figure \ref{fig:meanDPoverP}.
This plot shows that the reconstructed momentum is under-estimated at low momentum and over-estimated at high momentum.
Two effects can be the cause of such behavior :
\begin{enumerate}
\item The initial curvature comes from a global fit of all the CTD hits assuming that the tracks follow a helix.
This condition is only satisfied if the magnetic field is constant in all the CTD and if the particles does not loss any 
energy in the CTD. Since this is not completely the case, the curvature used as initial value of KFFIT is not 
the curvature at the inner wall of the CTD as it should be.
This bias is transmitted in KFFIT to the final value.
If we neglect energy loss in the CTD, this bias should be constant with momentum.
This bias on the initial curvature is responsible for the bias at high momentum.
\item At low momentum, the bias increases when the momentum decreases. This indicates an under-estimation of the energy loss.
This under-estimation can come from the approximations in the application of the Bethe Bloch formula 
(all materials are assumed to be Silicon and all particles are supposed to be pions) or can come from some 
dead material not taken into account in the code. 
So a simple factor 1.5 has been used firstly to multiply the Bethe Bloch formula.
But it has been found next that the beam pipe was never taken into account
in the algorithm (as seen in the next section). 
After correction of this bug, a factor 1.2 or 1.3 multiplying the Bethe Bloch formula had to be kept to correct the momentum bias. 
\end{enumerate}

It must be noted however that these corrections have been tuned on Monte Carlo samples.
It has been shown, by Hartmut Stadie and Vladimir Aushev\cite{HAR06}, that a remaining 
momentum shift was still there on real data with a value around 0.4\%. 
This effect is interpreted as the consequence of a slight mismatch between the magnetic field values 
used in the Monte Carlo reconstruction and the real one.
The final momentum is thus multiplied by a factor 1.0039 at the end of KFFIT before filling the ZTTRHL adamo tables.

\subsection{Correction for the material of the beam pipe}
In the release v1 and in all the previous releases, the beam pipe was represented by three ellipsoids with different thickness.
But this beam pipe had never been effective due to a missing parenthesis in the legacy KFFIT code.

The following condition had been used in the code to check whether the origin of the track is not included in the ellipsoid :
\begin{equation}
	if \left( ! \left( \left( \frac{2 x}{w} \right)^2 +  \left( \frac{2 y}{h} \right)^2 \right) \leq 1 \right)
	\label{eq:ineq}
\end{equation}
where $(x,y)$ is the axial position of the point of closest approach in the reference frame of the ellipsoid 
(where the center of the ellipsoid is the origin) and 
$w$ and $h$ are the width and the height of the ellipsoid.
If this condition is satisfied, the track does not cross the beam pipe.
So the calculation of multiple scattering or energy loss is not done.
But note that the exclamation mark (!), which in C means 
a reverse of the boolean value (true becomes false and vice versa), affects only the first term of the inequality and not the entire inequality. 
In consequence, the compiler converts the term after the exclamation point to boolean. 
And then it converts this boolean into an integer to calculate the inequality. 
This integer is either 0 or 1. But whatever the result, the condition will always be satisfied, meaning that the track does not cross the beam pipe.\\
So the condition has been simply corrected to :
\begin{equation}
	        if \left( ! \left( \left( \left( \frac{2 x}{w} \right)^2 +  \left( \frac{2 y}{h} \right)^2 \right) \leq 1 \right) \right)
		        \label{eq:ineq2}
\end{equation}

After this correction, other bugs appeared in some parts of the code that had apparently never been tested before.
All these problems have been corrected.\\

The consequences of the correction of this bug affects primarily the curvature.
The parameter used to correct the momentum bias (see previous section) could be reduced from 1.5 to 1.3.
Moreover the inclusion of the beam pipe affects the calculation of the covariance matrices because 
of multiple scattering.
This increases the covariance matrices by a factor around 1.2 which is consistent with the estimation of the pulls as shown in section \ref{sec:pull}.
		
\subsection{Algorithmic optimization and computing time}
Parts of the code have been re-written to optimize the computing time. 
KFFIT was indeed one of the largest contributions to the total latency of the data reprocessing. 
Here is a list of some improvements done amongst other things :
\begin{itemize}
	\item The geometry of the MVD had been loaded at each track. Since this is very time consuming, the MVD geometry is now loaded only once during the initialisation of KFFIT\footnote{That means that in the new version (v10) the alignment is not allowed to change within the same job!}.
	\item All the MVD clusters was loaded at each events. Now, only the needed MVD clusters (linked to a ZTPRHL track) are loaded.
	\item The simple methods, such as method returning a value, have now been inlined, so the compiler will replace the method call with a copy of the method body.
	\item All object arguments are now passed by constant references (const type\&) instead of objects.
	\item The operators designed for affectation (operator=) return now references instead of objects.
	\item Complex member objects of some classes were returned by some methods without being modified afterwards. As much as possible a constant reference is now returned.
	\item Some variables were calculated but used only for the pattern recognition mode of KFFIT which is not used in normal reconstruction. 
		These variables are now calculated only if this special mode is required.
\end{itemize}

\begin{figure}[htpb]
	\centering
	\includegraphics[height=5cm]{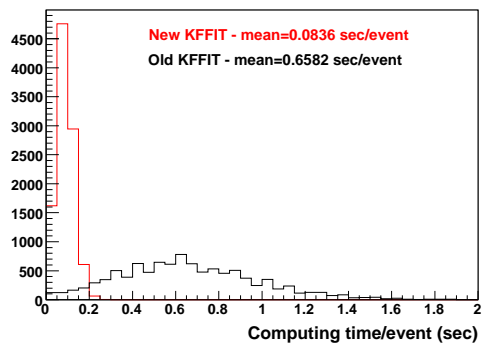}
	\caption{Mean processing time on an AMD opteron processor 252  (2.5 GHz).}
	\label{fig:timing}
\end{figure}

The figure \ref{fig:timing} shows the processing time per event on an AMD opteron processor 252 (2.5 GHz), 
in black before the improvements (release v1) and in red after (release v10).
The new KFFIT is 8.25 times faster than the old one.

An important structural improvement of the code is the use of standard C header files. 
KFFIT uses C structures describing ADAMO tables extensively.
In the old versions of KFFIT, local fortran headers were used and converted into C using cfortran.h.
This involved the tedious task of maintaining these local header files updated.
With transition to the centrally provided C headers, this maintenance is not needed anymore since the standard C headers are updated automatically for each releases.

\section{Performance}
\subsection{Sample used}
To test the new KFFIT, a sample of 100000 dijet MC events has been used.
The following track selection has been applied :
\begin{itemize}
	\item $p_T>0.2$ GeV
	\item $|\eta|<1.0$
	\item outer Super Layer $\geq 5$
	\item No forward MVD hits
\end{itemize}
The reconstructed tracks have been matched to the true tracks using the standard routine VMCU from the VCRECON package.

\subsection{Relative momentum resolution}
The various modifications of KFFIT improve primarily the curvature and thus the momentum resolution without affecting the other track parameters.
\begin{figure}[htpb]
	\centering
	\includegraphics[height=5cm]{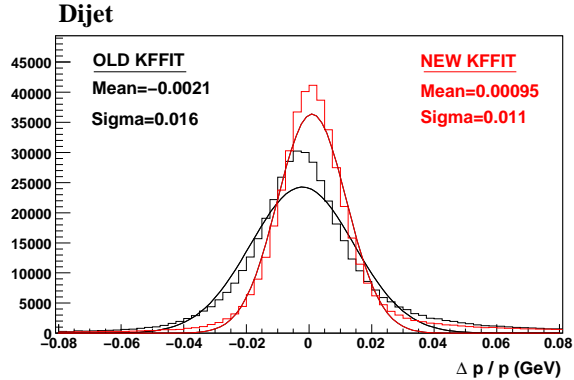}
	\caption{Relative momentum resolution for pions in a dijet sample. In black with the use of the old KFFIT (v1) and in red using the new release (v10).}
	\label{fig:DPOverP}
\end{figure}
The figure \ref{fig:DPOverP} shows the relative momentum resolution for pions in the dijet sample.
The black histogram corresponds to the resolution of the old version of KFFIT (release v1) and the red histograms corresponds to the new KFFIT (release v10).
An improvement of the width by a factor around 1.4 is observed.
\\
\begin{figure}[htpb]
	\centering
	\includegraphics[height=5cm]{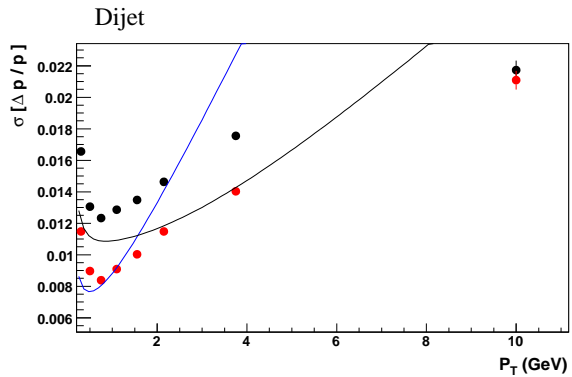}
	\caption{Relative momentum resolution as a function of the true tranverse momentum. The black points corresponds to the old KFFIT (v1) and the red points to the new KFFIT (v10). The black curve corresponds to the resolution obtained by Maddox \cite{MAD04} and the blue curve corresponds to the CTD resolution for HERA 1 \cite{HAL99}}
	\label{fig:RelMomentum}
\end{figure}
The figure \ref{fig:RelMomentum} has been obtained, using pions in the dijet sample, fitting vertical slices 
of $\Delta p_T/p_T$ by Gaussians and plotting the sigmas as a function of the transverse momentum. 
The black points corresponds to the old KFFIT (v1) and the red points to the new KFFIT. 
The comparison of these two distributions shows that the modifications of KFFIT improves the momentum in all the momentum ranges.
In addition, two curves have been added in the figure for a rough comparison.
The black curve comes from an expectation of the KFFIT momentum resolution obtained by E. Maddox in his 
thesis \cite{MAD04}, using a MC single muon sample which does not include any pattern recognition effects :
\begin{equation}
	\sigma p_T/p_T = 0.0026 p_T \oplus 0.0104 \oplus 0.0019/p_T
	\label{eq:maddox_res}
\end{equation}
This equation predicts a better momentum resolution than version v1 of KFFIT. This is because the samples used are different.
In the dijet sample, the pattern recognition can fail for busy events associating wrong hits to track.
Moreover, they are more type of particles in the dijet events with masses different from the mass of the pion.
This causes a wrong estimation of the energy loss in KFFIT.
For those reasons, the momentum resolutions contain long tails in the dijet sample and spreads the Gaussian used to fit the distributions.
But the new version of KFFIT gives an even better resolution than expected by equation \ref{eq:maddox_res} even using the dijet sample.
\\
The blue curve corresponds to the CTD momentum resolution obtained using samples of $D^*$ and semi-leptonic heavy quark decay events simulated in the HERA 1 architecture \cite{HAL99} :
\begin{equation}
        \sigma p_T/p_T = 0.0058 p_T \oplus 0.0065 \oplus 0.0014/p_T
        \label{eq:hal_res}
\end{equation}
This equation is a parametrisation of the momentum resolution for the CTD-only tracks when the MVD was not yet existing.
The predicted momentum resolution is very good at low momentum mainly because 
the amount of material was much smaller in the HERA 1 setup. 
With the inclusion of the MVD, multiple scattering in additional material causes a deterioration of the momentum resolution at low momentum.
But it is remarkable to see on the Fig. \ref{fig:RelMomentum} that the improved KFFIT is able to give a similar resolution at low momentum.

At high momentum, the curvatures of the tracks are small and the additional MVD hits increase the leverage used to calculate the track radius. 
In consequence the inclusion of the MVD improves considerably the momentum resolution at high momentum.
It must be noted however that the estimation of the momentum resolution presented here is always based on Monte Carlo samples assuming that the detector is perfectly aligned. 

\subsection{Pull} \label{sec:pull}
\begin{figure}[htpb]
	\centering
	\includegraphics[height=8cm]{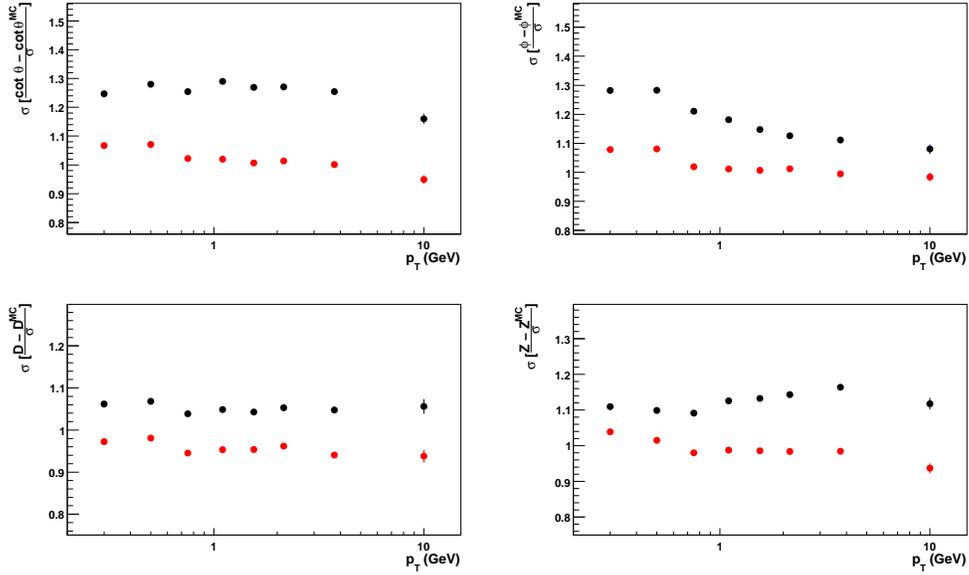}
	\caption{Standard deviation of the pull for the parameters $cot(\theta_{dip})$, $\phi_0$, $z_0$ and $D_0$. The black points correspond to the version v1 of KFFIT and the red points to the version v10.}
	\label{fig:pull}
\end{figure}
The figure \ref{fig:pull} shows the standard deviation of the pull for the parameters $cot(\theta_{dip})$, $\phi_0$, $z_0$ and $D_0$.
It appears that the new pulls are closer to 1 than in the previous version.
This is essentially due to the introduction of the beam pipe in KFFIT.
The beam pipe produces a non-negligible multiple scattering which increases the covariance matrix at the end of the Kalman Filter.
It must be noted nevertheless that the pulls shown on the figure \ref{fig:pull} have been obtained using the dijet MC sample
without any selection on the particle type. In this case, the pull distribution is not
Gaussian and contains long tails. 
So the resolution on parameters is slightly worse for other particles than pions. 
As a consequence, the errors are over-estimated for pions and under-estimated for particles with mass very far 
from the mass of the pions. But the average pull, for any samples and for all parameters, is in the range $1\pm0.2$.
\subsection{Momentum bias}
\begin{figure}[htbp]
  \centering
  \includegraphics[height=5cm]{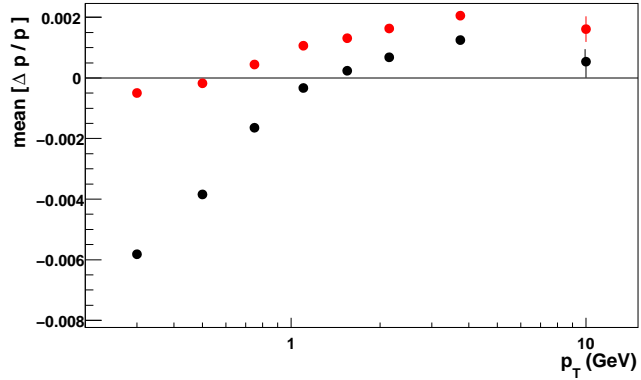}
  \caption{Mean $\Delta p/p$ as a function of the transverse momentum $p_T$.}
  \label{fig:MomBias}
\end{figure}

The figure \ref{fig:MomBias} shows the mean difference of true and reconstructed relative momentum as a function of the true transverse momentum.
The black points corresponds to the release 2005 (v1) and the red points corresponds to the latest version of KFFIT (v10).
It can be seen that the new distribution is more flat and in the range $\pm0.2$ \%.
This is essentially due to the introduction of the beam pipe in KFFIT in combination with the correction of the Bethe-Bloch formula.
Due to the principal limitation in the KFFIT material model that have been outlined before, no perfect compensation has been possible, but the achieved level of correction should clearly be good enough for most analyses.

\subsection{Test on 2005 data}
The new version of KFFIT has been tested on 2005 real data by Vladimir Aushev
on the $\Lambda$ and the $K_S^0$ mass distributions \cite{VLA06} and by Hartmut Stadie on the $J/\psi$ mass distribution \cite{HAR06b}.
\begin{figure}[htbp]
  \centering
  \includegraphics[height=10cm]{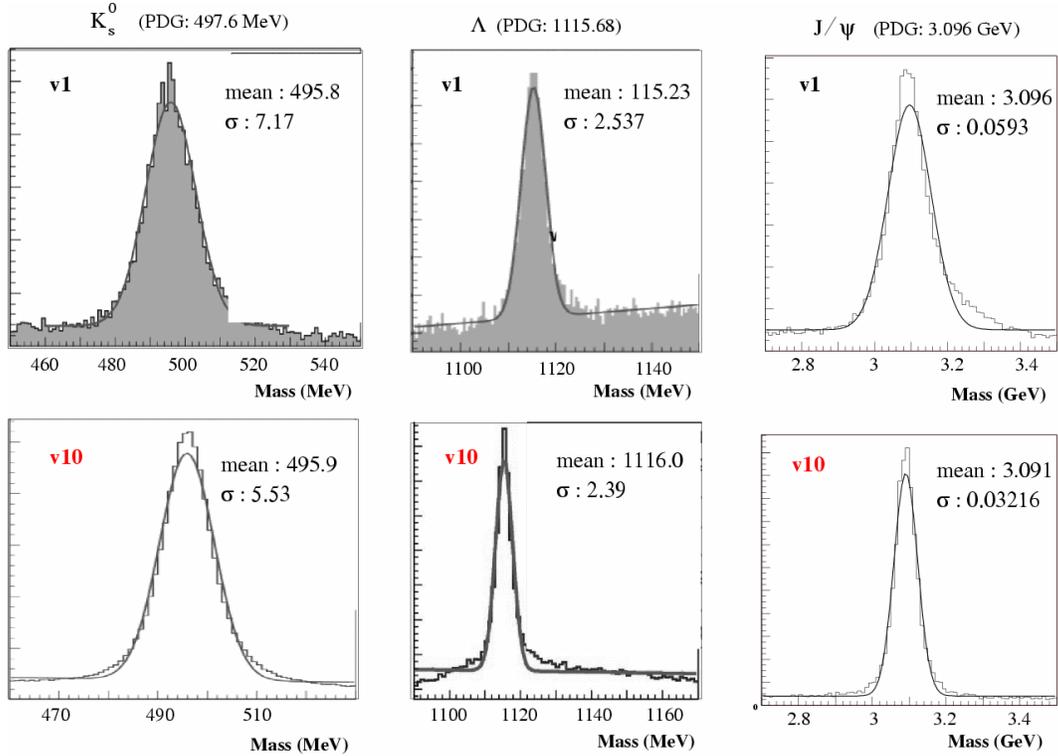}
  \caption{Mass distributions of the $K_S^0$, the $\Lambda$ and the $J/\psi$ for the version v1 of KFFIT (above) and for the new version v10 (below).}
  \label{fig:mass}
\end{figure}
As seen on figure \ref{fig:mass}, an improvement of about 7 \%, 22 \% and 45 \% have been observed respectively for the $\Lambda$, the $K_S^0$ and the $J/\psi$ mass resolution.
It can be seen that the mean value of the $J/\psi$ mass distribution is slightly worse for the new KFFIT and shifted by around 0.4 MeV.
This effect is not seen on the equivalent MC sample. So it has been interpreted as a consequence of a slight mismatch of the simulated and the real magnetic field.
As explained in section \ref{sec:bias} a factor of 1.0039\% has been applied to the final ZTTRHL momentum on real data to take into account this effect.
\section{Room for further possible development}
\begin{itemize}
\item To improve further the momentum resolution in KFFIT, the density and atomic numbers of dead material in the Bethe-Bloch formula need to be taken into account.
At present, all materials are assumed to be Silicon.
\item An other issue, would be to identify the particles (using dE/dx in the CTD or in the MVD) before entering in KFFIT and use its proper mass in the Bethe-Bloch formula.
\item The initial parameters used as input for KFFIT come from a global fit of all CTD hits assuming that the tracks can be described by one helix in all the CTD. 
This is an approximation. In order to start a proper Kalman Filter in the MVD we would need instead to use the track parameters at the inner wall of the CTD. 
\item At present tracks having STT hits are not fitted by VCRECON. So KFFIT uses ZTPRHL tracks as input which have not the proper 
parameters and covariance matrices to be used as input for the Kalman Filter. 
\item The contribution of the misalignment of the MVD to hit errors has to be determined properly. 
At present, the MVD hit intrinsic resolution $\sigma_{int}$ is increased by an extra term of 40 $\mu$m : $\sigma=\sqrt{\sigma _{int} + 40^2}$.
This is probably over-estimated since the new alignment constants have been improved using e-p collisions in addition to the cosmic alignment.
\item Finally, the association of MVD hits to tracks during the Kalman Filter could be be improved.
Indeed, a smoothing step could be added at the end of the Kalman Filter to remove MVD hits increasing the $\chi ^2$ of the fit abnormally ('outliers').
\end{itemize}

However, a new package called 'Rigorous Track Fit' is being written by Alexander Spiridinov. 
In this new algorithm, the Kalman Filter is performed not only on the MVD hits, but also on the CTD and on the STT hits.
Moreover, the fit takes into account the characteristic inhomogeneity of the magnetic field in the forward direction.
This algorithm shows already very promising results and will be used for tracks having STT hits for the next re-processing of the 2004 data.
So, maybe all the items mentioned above should be addressed by this Rigorous Track Fit approach.

\section*{Acknowledgments}
I would like to thank R. Mankel, G. Hartner, H. Stadie, V. Aushev and A. Spiridinov for useful discussions.
%
%

\newpage
\section*{Annex 1 : Preparation of output information}
This section describes the KFFIT output table : ZTTRHL.
The table \ref{tab:zttrhl} summarizes the various variables stored in this table.
\begin{table*}[htbp]
\begin{center}
\begin{tabular}{|l|l|l|} \hline
Name & Type & Description \\\hline
hel[5] & REAL & Helix parameters ($\phi_0$, $Q/R$, $QD_0$, $Z_0$, $\cot(\theta_dip)$)  \\
Cov[15] & REAL & Covariance matrix \\
Mom & REAL & Momentum \\
Chsq & REAL & The total $\chi^2$ (CTD+MVD) \\
Code & INTE & Code of the form NNJJII, where \\ 
     &      & NN=number of degrees of freedom (CTD+MVD),\\
     &      & JJ=outer SL, II=inner SL \\
StCrCsCz & INTE & aabbccdd: number used of \\
         &      & a=STT,b=CTDrp,c=CTDstereo,d=CTDz \\
BrBzWuWv & INTE & aabbccdd; number used of \\
         &      & a=BMVDrp,b=BMVDz,c=FMVDu,d=FMVDv \\
phiInner & REAL & Phi innermost \\
phiOuter & REAL & Phi outermost \\
mvddedx & REAL & Average of cluster sums (corrected) \\
elcorr & REAL & Strength of energy loss correction \\
method & INTE & Track fit used (1=kffit, 2=rtfit) \\
\hline
\end{tabular}
\end{center}
\caption{ZTTRHL ADAMO table for version v10 of KFFIT.}
\label{tab:zttrhl}
\end{table*}
\\
Some remarks have to be noticed :
\begin{itemize}
\item The track parameters, the covariance matrix and the momentum are calculated at the point of closest approach to the origin.
\item The momentum ({\bf Mom}) is corrected using real data : $Mom = Mom*1.0039$
\item The {\bf Chsq} is calculated as the sum of the CTD and the MVD $\chi^2$ plus an additional term which takes into account 
the fact that the inclusion of the MVD hits will also change the trajectory within the CTD, which should give rise to an 
additional $\chi^2$ contribution :
\begin{equation}
  \chi^2=\chi^2_{CTD}+\chi^2_{MVD}+\Delta \vec{a}^{T} \boldsymbol{C}_{CTD}^{-1} \Delta \vec{a}
  \label{eq:chi2}
\end{equation}
where $\Delta \vec{a}^{T}=\vec{a}_{CTD}-\vec{a}_{MVD}$, the difference between the CTD track parameters and the KFFIT track parameters ($\vec{a}_{MVD}=hel[5]$).
$\boldsymbol{C}_{CTD}$ is the covariance matrix of the CTD tracks.
\item The number of degrees of freedom stored in {\bf Code} is equal to the number of degree of freedom from the CTD input tracks plus the number of MVD hits used in KFFIT.
\item {\bf StCrCsCz} does not involve MVD information and is just a copy from the ZTPRHL table.
\item {\bf phiInner} is the $\phi$ angle of the track at the inner most hit. If this hit is in the MVD, it is calculated as the $\phi$ angle of the track at the intersection point of this track with the sensor of this hit.
\item {\bf phiOuter} is the $\phi$ angle of the track at the outer most hit and is calculated from the ZTPRHL table :
\begin{eqnarray}
  s_{outer}^{ztprhl}=(\phi_{outer}^{ztprhl}-\phi_0^{ztprhl})/(QR)^{ztprhl} \\
  \phi_{outer}^{zttrhl} = \phi_0^{zttrhl}-(QR)^{zttrhl}s_{outer}^{zttrhl}
  \label{eq:phiouter}
\end{eqnarray}
\item {\bf mvddedx} is the average of the corrected cluster sums :
\begin{equation}
  mvddedx=\sum_{i=0}^{N} \frac{ Q_i \cos(\theta_i)}{N} 
  \label{eq:mvddedx}
\end{equation}  
where $N$ is the number of MVD hits fitted in KFFIT, $Q_i$ is the collected charge for the $i$\_th hit (from MVRECC\_signal) and $\theta_i$ is the angle of the track with respect to the normal of the sensor of this hit.
\item {\bf elcorr} is the total momentum loss in the MVD (in GeV). This value can be used in a further analysis to correct any remaining momentum bias due to approximations in the Bethe Bloch formula (mass of the particles=mass of the pion, all material is Silicon,...).
\end{itemize}

\section*{Annex 2: History of KFFIT}
The initial version of KFFIT used in this study was v2005b.1 which differs slightly from v2005a.1 because contains some speed-up implemented by Hartmut Stadie.
\subsection*{v1 = v2006a - February 22 2006}
\begin{itemize}
\item Version equivalent than v2005b.1 but with headers modified to use the standard c headers stored in :
\verb+$ZSYS_ROOT/Released/zeus/ZeusAdamo/$ZEUSRELEASE/inc/windows\_c+
\item call to cfortran.h removed.
\end{itemize}
\subsection*{v2 - March 13 2006}
\begin{itemize}
\item Input tracks are constructed in the class 'Event' and not anymore in the main class 'Fit'.
\end{itemize}
\subsection*{v3 - March 17 2006}
\begin{itemize}
\item MVD hits are now attached to input tracks in Event
\item All material are now defined in kfinit and not anymore for each tracks
\item Various variables calculated under the condition 'if (mycuts.refitPR())' corresponding to a special mode of KFRECON : mode Pattern Recognition.
\item CTD only tracks (obtained from the VCTRHL adamo table filled by VCGTRGASIS) are used as input of KFFIT if the number of MVD hits $>$ 0 and if CTDouter$\geq$3.
\item Subroutine VCUNDOSCAT is used to remove multiple scattering in the CTD only input tracks.
\item New methods created to test the new CTD only input tracks.
\item Input covariance matrices rescaled to take into account the oversizing of the CTD hits in VCRECON.
\end{itemize}
\subsection*{v4 - June 10 2006}
\begin{itemize}
\item Intersection with sensor not anymore required to accept an MVD hit.
\item MVD hits sorted in the constructor of the class 'Event'.
\item $\Delta \chi^2$ cut increased to 1000.
\item Possibility to use either CTD only tracks as input or ZTPRHL as in version v1 using the control card : KFFIT-UsePR.
\item Variable name ThroughMVD replaced by thrMVD.
\end{itemize}
\subsection*{v5 = v2006a.1 - July 05 2006 }
\begin{itemize}
\item MVD hits resolution multiplied by 1.3 (needed to get output pulls close to 1 at high momentum).
\item New method to calculate residuals and error on parameters.
\end{itemize}
\subsection*{v6 = v2006a.2 = v2006a.2.b - August 31 2006}
\begin{itemize}
\item Energy loss multiplied by 1.5 to correct momentum bias.
\item New method to calculate energy straggling (in test)
\item Check on the output covariance matrices : if not positive redo the Kalman Filter with ZTPRHL tracks as input.
\end{itemize}
\subsection*{v7 - September 21 2006}
\begin{itemize}
\item Test on input CTD only covariance matrices : if not positive start iteration on VCUNDOSCAT increasing at each step the initial momentum used.
\item Magnetic field is now depending on the position and not anymore a constant value.
This influence calculation of energy loss in material.
\item The factor used to multiply energy loss using Bethe Block formula to take into
account missing material is becoming a new KFFIT card : KFFIT-dEdxCorr
\item Speed-up of the code : part have been re-written.
Important modification : All the MVD hits are not loaded anymore, but only the one attached to a ZTPRHL track.
\end{itemize}
\subsection*{v8 - September 29 2006}
\begin{itemize}
\item Bug corrected : the beam pipe was never taken into account.
\item Default value of dEdxCorr set to 1.2 and not anymore 1.5.
\item 1/R pull corrected by addition of 'error due to energy loss approximation'
and energy straggling.
\end{itemize}
\subsection*{v9 - October 24 2006}
\begin{itemize}
\item Initialisation removed from orangeinit and only done in KFFIT.
\item Default value of dEdxCorr set to 1.3
\item Momentum correction at the end of KFFIT for real data :
   zttrhl\_.Mom = zttrhl\_.Mom*1.0039
\item Fill 2 new variables in ZTTRHL table : method=1 for KFFIT and elcorr=total energy loss by a pion in the MVD.
\item Fill 1 new variable in ztrprm table : Weight=1
\end{itemize}
\subsection*{v10pre = v2006a.3 - 5 December 2006}
\begin{itemize}
\item Variable zttrhl\_.StCrCsCz is now just a copy of ZTPRHL
\item The fit (or calculation of energy loss,...) is not done anymore for CTD standalone
tracks but the ZTPRHL tracks are simply copied to ZTTRHL.
\item zttrhl\_.mvddedx = Average charge sum corrected by cos($\theta$) where $\theta$
is the angle of the track direction at the intersection point with respect to the 
normal to the sensor.
\item zttrhl\_.elcorr = Sum of the energy loss in all MVD material.
\end{itemize}
\subsection*{v10 = v2006a.4 - 11 December 2006}
\begin{itemize}
\item The fit is not done anymore for tracks with ztprhl\_VCTRHL==INULL
\item zttrhl\_.phiOuter and zttrhl\_.phiInner are now correctly filled
\item $\chi^2=\chi^2_{mvd} + \chi^2_{ctd} + f$ where f is a factor to take into account the increase of $\chi^2_{ctd}$ after modifying the tracks in KFFIT.
\item $Ndf=Ndf_{ctd}+Ndf_{mvd}$ where $Ndf$ is the number of degree of freedom stored in zttrhl\_.Code.
\end{itemize}
\end{document}